\documentclass[onecolumn,aps,pre,groupedaddress,showpacs,floatfix]{revtex4-1}
\usepackage{graphicx}
\usepackage{epsfig}
\usepackage{amsfonts}
\usepackage{color}
\usepackage{ulem}
\usepackage{amsmath}
\usepackage{mathrsfs}
\usepackage{subfigure}
\usepackage[autostyle]{csquotes}

\begin{document}

\title{Double diffusivity model under stochastic forcing}

\author{Amit K Chattopadhyay}
\affiliation{                    
Mathematics and
Aston Institute of Materials Research (AMRI), Aston University, Aston Triangle, Birmingham, B4 7ET, United Kingdom}
\email{a.k.chattopadhyay@aston.ac.uk}
\author{Elias C. Aifantis}
\affiliation{Laboratory of Mechanics and Materials, Aristotle University of Thessaloniki, GR-54124 Thessaloniki, Greece \\
Michigan Technological University, Houghton Michigan 49931, USA,  \\ ITMO University, St. Petersburg 197101, Russia and BUCEA, Beijing 100044, China}
\email{mom@mom.gen.auth.gr}

\begin{abstract}
\noindent
The \enquote{double diffusivity} model was proposed in the late 1970s, and reworked in the early 1980s, as a continuum counterpart to existing discrete models of diffusion corresponding to high diffusivity paths, such as grain boundaries and dislocation lines. It was later rejuvenated in the 1990s to interpret experimental results on diffusion in polycrystalline {{and}} nanocrystalline specimens where grain boundaries {{and}} triple grain boundary junctions act as high diffusivity paths. Technically, the model pans out as a system of coupled {\it Fick type} diffusion equations to represent \enquote{regular} and \enquote{high} diffusivity paths with \enquote{source terms} accounting for the mass exchange between the two paths. The model remit was extended by analogy to describe flow in porous media with double porosity, as well as to model heat conduction in media with two non-equilibrium local temperature baths e.g. ion and electron baths. Uncoupling of the two partial differential equations leads to a higher-ordered diffusion equation, solutions of which could be obtained in terms of clasical diffusion equation solutions. Similar equations could also be derived within an \enquote{internal length} gradient (ILG) mechanics formulation applied to diffusion problems, {\it i.e.}, by introducing nonlocal effects, together with inertia and viscosity, in a mechanics based formulation of diffusion theory. While being remarkably successful in studies related to various aspects of transport in inhomogeneous media with deterministic microstructures and nanostructures, its implications in the presence of stochasticity have not yet been considered. This issue becomes particularly important in the case of diffusion in nanopolycrystals whose deterministic ILG based theoretical calculations predict a relaxation time that is only about one-tenth of the actual experimentally verified timescale. This article provides the \enquote{missing link} in this estimation by adding a vital element in the ILG structure, that of stochasticity, {{that takes into account all boundary layer fluctuations}}. Our stochastic-ILG diffusion calculation confirms rapprochement between theory and experiment, thereby benchmarking a new generation of gradient-based continuum models that conform closer to real life fluctuating environments. 
\end{abstract}
\date{\today}


\maketitle

\newpage

\section{Introduction}

\noindent
The subject of material science has traditionally dealt with hard and soft matter based objects, typically at observable macroscopic, and mesoscopic, and microscopic scales. Until very recently, most branches of traditional science used to be expressible within such length scales for which appropriate phenomenological theories have been well established over time. Atomistic or lattice based approaches have also been developed and related to molecular dynamics {{and}} quantum mechanical simulation codes have been advanced. However, the regime between the microscopic and atomic scale, {\it i.e.}, the nanoscale regime, especially the length scale between 5 and 100 nm, still remains a major modeling challenge. 

In particular, the advent of nanocrystalline and ultrafine grain nanocomposite materials, along with their tremendous possibilities in material engineering implementations have challenged all existing conventional and well established theoretical and experimental realizations. The cause of this can be easily attributed to the fact that at manometric scales, most materials, both soft and hard, often have widely different physicomechanical and thermochemical properties to their macroscopic counterparts that prove elusive to existing scientific analyses. 

A simple case in hand is that of diffusion in nanocrystals, and more generally, in nanocomposites, for which it has been shown that the diffusivity could be many orders of magnitude larger than the diffusion constant for bulk lattice diffusion of mesoscopic and microscopic scale materials \cite{Aifantis2011}. This is in line with other mechanical, electrical, magnetic and chemical properties of nanoscopic objects where nanoscale moduli differ significantly from their macroscale and microscale equivalents. A compromise between the atomistic and conventional continuum mechanical engineering modeling approaches for nanopolycrystals is reviewed in \cite{Aifantis2011}, within a Laplacian based internal length gradient (ILG) generalization of classical deformation (elasticity, plasticity) and diffusion theories. A more thorough analysis, including size effects in chemomechanics and electromechanics, as well as the role of stochasticity, due to internal stress fluctuations, can be availed in detail from \cite{Aifantis2016}. This article refrains from explicit discussion of the role of stochasticity in diffusion at nanoscales, a task earmarked for the present article. 

Nanoscale diffusion or diffusion in nanopolycrystals in particular, is viewed as a two-phase process, one through the bulk and the other across the grain boundary (GB) space. In the case representing an admixture of two families of grain boundaries (low and high GBs), depending on the proportion of the two types of GBs, it may be necessary to account for a third type of diffusion, a variant of the paradigmatic stochastic nanodiffusion \cite{akc-eca-2016} dynamics. For decreasing grain size, the density of the triple grain boundary junction (TJ) increases significantly and then, TJ has also to be identified as a separate family of high-diffusivity paths. In any case, independently of how one will identify the types of the families of paths available for diffusion, the mathematical model that we discuss here distinguishes between only two types of paths, fast and slow. These two paths with varying rate kinetics are differentiated by assigning two different diffusivities, thereby allowing mass transfer of diffusion species between them. For example, in the case of very small grain sizes ($\sim$10 nm), where a large number of triple junctions is present, the two types of diffusion may be identified with GB and TJ, while bulk diffusion may be neglected as being comparably at a much slower rate. For severe plastic deformation (SPD) fabricated polycrystals with larger grain sizes ($\sim$100 nm), the two types of high diffusivity paths may be identified with low-angle (equilibrium) and high-angle (nonequilibrium) grain boundaries which are densely populated with dislocation/disclination defects. For microscale polycrystals with even larger grain sizes ($\sim$1 $\mu$m), the two families of diffusion paths are represented by the bulk/grain interior and the surrounding/grain boundary space. With the above interpretation, we focus on modeling stochastically enhanced counterpart of the deterministic continuum model proposed and analyzed in \cite{Aifantis1979JApp, Aifantis1979Acta, Aifantis1980Quarter, Aifantis1980Acta, Aifantis1980Engg, Beskos1980, Aifantis1980Mech, Konstantinidis1998,Konstantinidis1999, Konstantinidis2001, Forest2010, Kalampakas2012}.

More specifically, we elaborate on the coupled system of the two partial differential equations proposed in \cite{Aifantis1979JApp}, solved in \cite{Aifantis1980Quarter}, and thereafter extended in \cite{Aifantis1980Acta, Aifantis1980Engg} to explain double porosity in media and through \cite{Beskos1980, Aifantis1980Mech} to describe heat transport in materials with two temperatures. All of these theoretical enclaves were eventually implemented in \cite{Konstantinidis1998, Konstantinidis1999, Konstantinidis2001} to interpret experimental measurements of diffusion in polycrystals and nanocrystalline aggregates. Some thermodynamic aspects of the deterministic double diffusivity model have been relatively recently discussed in \cite{Forest2010}, where the approach is probability based. Phenomenological coefficients have recently been provided in \cite{Kalampakas2012}. 

{{From the above discussion, it follows that a vital contributing factor that has remained unattended in the legion of double diffusivity studies is that of thermally induced fluctuations arising out of the structural difference between lattice and grain boundary spacings, especially close to the boundary layers as also arising out of material imperfections and structural randomness. In line with the well established legion of stochastically forced flow models \cite{Forster1977, DeDominicis1979, Yakhot1986, Chattopadhyay2000} representing archaic dynamical randomness generated close to the sheared boundary layers, as also due to structural imperfections, like crack propagation, this article will explore this realistic limit of double-diffusion, thereby accounting for all modes of randomness. This will be structured within the well-knit Langevin formulation of stochastic dynamics \cite{Risken, Barabasi}}}. Phenomenologically, this can be seen as an external stochastic force that is randomly redistributing the relevant spatial structure, for example, that of the high-diffusivity paths, in which a stochastic increase (decrease) in temperature extends (contracts) the interlayer grain boundary distance between two nanosized grains and thus alters the configuration of the structural defects (dislocations, disclinations). Such a multi ensembled stochastic reorganization of the lattice space distribution is sometimes known to create new universality classes \cite{Barabasi, Cross1993} as well.

\section{The Models}
\label{section_Model}

\noindent
Crack-fracture propagation through wavefronts and diffusion in complex media have long been known to be complementary physical realizations that can be addressed using a combination of continuum mechanics and numerical simulation \cite{Barabasi}. ILG, on the other hand, has been established as a powerful theoretical tool to address stress-strain deformation aspects in materials, as well as in related thermomechanical, electromechanical and chemomechanical processes as also in complex heterogeneous media \cite{Aifantis2011, Aifantis2016, Aifantis2014}. The role of randomness and some combined deterministic gradient-stochastic models have been considered in \cite{Barabasi, Aifantis2016, Aifantis2014} for \enquote{higher-ordered} deformation models. In contrast, the role of stochasticity has not been considered for higher-ordered diffusion models such that the double-diffusivity equations or the ${4}^{\text{th}}$ ordered diffusion equations resulting from their uncoupling, following on from the ILG mechanics.

Our theoretical structure will combine two models separately in which Model 1 will define the spatiotemporal dynamics of each individual phase with respect to the concentration of the diffusing species in the phase concerned. With reference to \cite{Aifantis2011}, where the deterministic version of this model referred to bivariate dynamics, we will call this the \enquote{non-conserved double-diffusive model} (NDD). Model 2, on the other hand, will focus on the higher-ordered diffusion equations resulting from uncoupling the NDD equations which also hold for the total concentration, {\it i.e.}, the sum of the two concentrations in the \enquote{slow} and \enquote{high} diffusivity paths. 
Once again, keeping in mind the origin of its deterministic analog, we will call this the \enquote{conserved double-diffusive model}. 

\noindent
A key outcome of this analysis is the fundamentally different dynamical structure functions (two-point spatiotemporal functions) estimated for these two classes of models. This is highly non-trivial, and a consequence of symmetry violation due to the stochastic forcing term in these models, since the deterministic versions of these two classes of models converged to the same \enquote{universality class}. 

\subsection{Forced non-conserved double-diffusion (NDD) model}
In line with the narrative presented earlier, we will arrive at the first {\it stochastic gradient nanomechanics} (SGNM) model of nanodiffusion starting from a deterministic Fick's diffusion like model (Model 1, as referred to in \cite{Aifantis2011}). Defining $\rho_1$ and $\rho_2$ as the concentrations of the diffusing species, for example, in the intercrystalline (IC) and TJ space, respectively, the stochastically forced Fick's diffusion model can be written as

\begin{subequations}
\begin{equation}
\frac{\partial \rho_1}{\partial t} = D_1 \nabla^2 \rho_1 - \kappa_1 \rho_1 + \kappa_2 \rho_2 + \eta_1({\bf x},t) 
\label{modeleqna}
\end{equation}
\begin{equation}
\frac{\partial \rho_2}{\partial t} = D_2 \nabla^2 \rho_2 + \kappa_1 \rho_1 - \kappa_2 \rho_2 + \eta_2({\bf x},t).
\label{modeleqnb}
\end{equation}
\end{subequations}

\noindent
In Eqs. (\ref{modeleqna}) and (\ref{modeleqnb}) above, $D_1$ and $D_2$ refer to the diffusion constants in the IC and TJ phases respectively, with $\kappa_1$ and $\kappa_2$ denoting the respective formation or depletion rate of diffusive substances at these phases. A negative sign before $\kappa_i\:\rho_i$ ($i$=1, 2) would indicate depletion, {\it i.e.}, loss of diffusion species \enquote{jumping} to the other \enquote{phase} while a positive sign would indicate replenishment of that particular phase. In the given form above, the deterministic concentration abides a certain conservation law (detailed below) with $\eta_i({\bf x},t)$ (i=1,2) being the white thermal noise perturbing this dynamics and is defined as follows

\begin{subequations}
\begin{eqnarray}
\langle \eta_i({\bf x},t) \rangle &=& 0 \\
\langle \eta_i({\bf x},t)\:\eta_j({\bf x'},t') \rangle &=& 2\delta_{ij}\:\gamma_i\:\delta^d({\bf x-\bf x'}) \delta(t-t'),
\label{noiseeqn}
\end{eqnarray}
\end{subequations}

\noindent
where $d$ represents the spatial dimension and $\gamma_i$ ($i$=1,2) are the noise strengths corresponding to $\eta_1$ and $\eta_2$ respectively.

\subsection{Forced conserved couble-diffusion (CDD) model}

\noindent
{{Using a series of transformations, $\tau=\dfrac{1}{\kappa_1 + \kappa_2}$, $D=\tau(\kappa_1 D_2 + \kappa_2 D_1)$, $c^* = \tau(D_1+D_2)/D$ and $c=-\tau D_1 D_2/D$, as detailed in \cite{Aifantis2011}, the deterministic model corresponding to Eqs. (\ref{modeleqna}) and (\ref{modeleqnb}) (for the noiseless case $\eta_i=0$)}} can be easily shown to abide by a conservation law in a scaled variable set defined as follows:

\begin{equation}
\partial_t \rho + \tau {\partial_t}^2 \rho = D \nabla^2 \rho + c^{*} D \partial_t\:\nabla^2 \rho + c D\:\nabla^4 \rho.
\label{scaledmodeleqn1}
\end{equation}

\noindent
This fourth-order equation containing, in addition to the classical Fick's law, a second time derivative inertial or \enquote{telegrapher} term ($\tau \nabla^2 \rho$); a third order mixed spatiotemporal or pseudoparabolic term ($c^* D \partial_t \nabla^2 \rho$), and a fourth order spatial or biharmonic term ($c D \nabla^4 \rho$) is obtained by uncoupling Eqs. (\ref{modeleqna}) and (\ref{modeleqnb}) with $\eta_2=0$. It holds for both individual concentrations $\rho_1$ and $\rho_2$, as well as for its sum $\rho=\rho_1+\rho_2$, {\it i.e.}, the total concentration.
{{The boundary layer fluctuation and random structural imperfection perturbed stochastic model can be derived from the deterministic model defined in Eq. (\ref{scaledmodeleqn1}) as a stochastically forced model with an additive (uncorrelated white) noise. The corresponding root-mean-square spatiotemporal \enquote{width} will define the spatiotemporal evolution of the interface separating the two phases. The resultant model (Model 2, as referred to in \cite{Aifantis2011}) is defined below}}

\begin{equation}
\partial_t \rho + \tau {\partial_t}^2 \rho = D \nabla^2 \rho + c^{*} D \partial_t\:\nabla^2 \rho + c D\:\nabla^4 \rho + \eta({\bf x},t),
\label{scaledmodeleqn}
\end{equation}

\noindent
where $\rho({\bf x},t)$ is the mass density of the separation width of the two phases,
in which $\tau = \frac{1}{\kappa_1 + \kappa_2}, \:
D = \tau(\kappa_1 D_2 + \kappa_2 D_1),\:
c^{*} = \tau\frac{(D_1+D_2)}{D}, \:
c = -\tau \frac{D_1 D_2}{D}$, and $\eta({\bf x},t)$ is the stochastic fluctuation.

For the special case for which $D_1=D_2=D$ in Eqs. (\ref{modeleqna}) and (\ref{modeleqnb}) give 

\begin{equation}
\partial_t \rho = D \nabla^2 \rho,
\label{conservationeqn}
\end{equation}

\noindent
that is a simple diffusion equation in the variable $\rho=\rho_1+\rho_2$. One must note that both spatiotemporal and reflection conservation as inherent to Eq. (\ref{scaledmodeleqn}) will be lost in the presence of a non-zero noise which is our starting model. {{Instances of such general applications of double-diffusion model in analyzing heterogeneous growth process have also been explored \cite{Showalter2004, Klein2012}.}} \\

\par
\noindent
In the following sections, we will separately analyze the single phase dynamics of the respective IC and TJ concentrations (Model 1) by estimating autocorrelation functions of the variables $\rho_1$ and $\rho_2$ respectively, from Eqs. (\ref{modeleqna}) and (\ref{modeleqnb}). These individual autocorrelations will then be compared against the autocorrelation evaluated from Eq. (\ref{scaledmodeleqn}) that defines the dynamics of the total concentration. The following sections will then estimate the spatiotemporal dynamics of the interface in presence of noise from estimation of the spatiotemporal correlation functions. This later part of the analysis will be compared against separate \enquote{thin} and \enquote{thick}-film conditions as detailed in \cite{Aifantis2011} to establish the importance of the stochastic contribution.  

\section{Phase Evolution Dynamics of the NDD Model}

\noindent
The primary focus in this section will be the evaluation of the individual phase dynamics of the IC and TJ phase variables $\rho_1$ and $\rho_2$ of the NDD model. Our starting point here will be the Fourier transformation of the core model presented in Eqs. (\ref{modeleqna}) and (\ref{modeleqnb}), defined by variables $({\bf k},\omega)$ of Eqs. (\ref{modeleqna}) and (\ref{modeleqnb}) to arrive at the following matrix form

\begin{equation}
\mathcal{M}
\left( \begin{array}{c}
\hat \rho_1 \\
\hat \rho_2 \\
\end{array} \right) =
\left( \begin{array}{c}
\hat \eta_1 \\
\hat \eta_2 \\
\end{array} \right),
\label{matrixeqn1}
\end{equation}

where the matrix $\mathcal{M}$ is defined as 

\begin{equation*}
\mathcal{M}=\left( \begin{array}{cc}
-i \omega+D_1 k^2 + \kappa_1 & -\kappa_2 \\
-\kappa_1 & -i\omega +D_2 k^2 +\kappa_2 \\
\end{array} \right).
\end{equation*} 

\noindent
Here $\hat \rho_1$ and $\hat \rho_2$ represent the Fourier transformed version of the $(\rho_1,\rho_2)$ variables in the $({\bf k},\omega)$ space (3+1 dimensional) while the Fourier transforms themselves abide by the following generic form

\begin{equation}
\psi({\bf x},t) = \int\:d^d{\bf k}\:\int\:d{\omega}\:{\hat \psi}({\bf k},\omega)\:e^{(i{\bf k.x}-\omega t)},
\label{fouriereqn}
\end{equation}

\noindent
where $\psi$ generically represents either $\rho_i$ or $\eta_i$.
From Eq. (\ref{matrixeqn1}), we get

\begin{equation}
\left( \begin{array}{c}
\hat \rho_1 \\
\hat \rho_2 \\
\end{array} \right)=\mathcal{M}^{-1} \left( \begin{array}{c}
\hat \eta_1 \\
\hat \eta_2 \\
\end{array} \right).
\label{matrixeqn2}
\end{equation}

\vspace{0.2cm}
\noindent
This inverse of the matrix $\mathcal{M}$ takes the form 

 \begin{equation*}
\mathcal{M}^{-1} = \left( \begin{array}{cc}
M_1 & M_2 \\
M_3 & M_4 \\
\end{array} \right),
\end{equation*}

\noindent
where 
\begin{eqnarray*}
M_1 &=& \dfrac{-i \omega^2 + k^2 D_2 + \kappa_2}{-\kappa_1 \kappa_2 + (-i \omega^2 + k^2 D_1 + \kappa_1)(-i \omega^2+k^2 D_2S+\kappa_2)}, \\
M_2 &=& \dfrac{\kappa_2}{-\kappa_1 \kappa_2 + (-i \omega^2 + k^2 D_1 + \kappa_1)(-i \omega^2+k^2 D_2S+\kappa_2)}, \\
M_3 &=& \dfrac{\kappa_1}{-\kappa_1 \kappa_2 + (-i \omega^2 + k^2 D_1 + \kappa_1)(-i \omega^2+k^2 D_2S+\kappa_2)}, \\
M_4 &=& \dfrac{-i \omega^2 + k^2 D_1 + \kappa_1}{-\kappa_1 \kappa_2 + (-i \omega^2 + k^2 D_1 + \kappa_1)(-i \omega^2+k^2 D_2S+\kappa_2)}.\\
\end{eqnarray*}

As like in any stochastically driven model \cite{DeDominicis1979, Chattopadhyay2000}, in the stochastically ensemble averaged state, the measurables will be the (Brownian) root-mean-square averaged quantities of their deterministic equivalents. The corresponding mean energy dissipation rate in such an ensemble-averaged state is thus defined as the time averaged kinetic term $\frac{1}{2}{\langle \partial_t \rho_i({\bf x},t)*\partial_t \rho_i({\bf x},t) \rangle}$ ($i$=1, 2) that necessitates evaluation of the following two autocorrelation functions:

\begin{subequations}
\begin{eqnarray}
{\rho_1}^{\text{rms}} &=& \sqrt{\langle \rho_1^2({\bf x},t) \rangle} \label{autocorrelations1a}\\
{\rho_2}^{\text{rms}} &=& \sqrt{\langle \rho_2^2({\bf x},t) \rangle} \label{autocorrelations1b}
\end{eqnarray}
\end{subequations}

and a complementary set of cross-correlation functions given by

\begin{eqnarray}
{\rho_{12}}^{\text{rms}} &=& \sqrt{\langle {\rho_1}^{*}({\bf x},t) \rho_2({\bf x},t) \rangle}=\sqrt{\langle {\rho_1}({\bf x},t)\rho_2^*({\bf x},t) \rangle}.
\label{crosscorrelations1}
\end{eqnarray}

In the above, the superscript \enquote{rms} stands for the root-mean-squares of the respective quantities under consideration post the stochastic (Brownian) average that is indicated by the \enquote{$< . >$} sign, while $\rho_i^*$ is the complex conjugate of $\rho_i$. The cross-correlations of the quantities with their complex conjugates emphasize the importance of the attenuation term in the dynamics (complex quantity); the fluctuation-dissipation theorem is always implicitly assumed in such analyses. In this work, we will assume noise cross-correlation to be zero (that is uncorrelated) and hence $\rho_{12}^{\text{rms}}=0$. The rms quantities are the ones of our interest, as this is what an experimental measurement will see, an allusion to the classical Brownian dynamics \cite{Barabasi}.

Eqs. (\ref{autocorrelations1a}) and (\ref{autocorrelations1b}) can be explicitly written as 

\begin{eqnarray}
{({\rho_1}^{\text{rms}})}^2 &=& 
\int d^d{\bf k}\:\int d{\omega}\:\langle \rho_1({\bf k},\omega)\:{\rho_1}^{*}({\bf -k},-\omega) \rangle, \nonumber \\
{({\rho_2}^{\text{rms}})}^2 &=& 
 \int d^d{\bf k}\:\int d{\omega}\:\langle \rho_2({\bf k},\omega)\:{\rho_2}^{*}(-{\bf k},-\omega) \rangle.
\label{autocorrelations2}
\end{eqnarray}

\subsection{Phase autocorrelation and crosscorrelation}

Equation (\ref{matrixeqn2}) can be solved to obtain the autocorrelation functions in the $k-\omega$ space:

\begin{subequations}
\begin{eqnarray}
\langle \rho_1({\bf k},\omega)\:{\rho_1}^{*}(-{\bf k},-\omega) \rangle
&=& \frac{2\gamma_1 (\omega^4+{(D_2 k^2 + \kappa_2)}^2) + 2\gamma_2 \kappa_2^2}{{{\zeta_n}}}, \\
 \langle \rho_2({\bf k},\omega)\:{\rho_2}^{*}(-{\bf k},-\omega) \rangle
&=& \frac{2\gamma_2 (\omega^4+{(D_1 k^2 + \kappa_1)}^2) + 2\gamma_1 \kappa_1^2}{{{\zeta_n}}},
\label{autocorr}
\end{eqnarray}
\end{subequations}

\noindent
where the quantity ${{\zeta_n}}=(\omega^4+ D_2^2 k^4)[\omega^4+{(D_1 k^2 + \kappa_1)}^2]+2[\kappa_1 \omega^4 + D_2 k^2(\omega^4 + D_1 k^2(D_1 k^2 + \kappa_1))]\kappa_2 + (\omega^4+D_1^2 k^4) \kappa_2^2$ defines the pole structure and hence possible discontinuities in the spectral dynamics. 

The Fourier transformed cross-correlation is even more interesting in that it shows a complex form in which the complex part represents attenuation. The structure looks as follows
\begin{equation}
<\rho_1({\bf k},\omega)\:{\rho_2}^{*}(-{\bf k},-\omega)> = 
\bigg[\frac{2\gamma_1 \kappa_1 (D_2 k^2 + \kappa_2) + 2\gamma_2 \kappa_2(D_1 k^2 + \kappa_1)}{\zeta_n}\bigg] +  i\bigg[\frac{2\gamma_1 \omega^2 \kappa_1 + 2\gamma_2 \omega^2 \kappa_2}{{\zeta_n}}\bigg].
\label{crosscorr}
\end{equation}

\begin{center}
\begin{figure}[tbp]
\includegraphics[height=10.0cm,width=12.0cm]{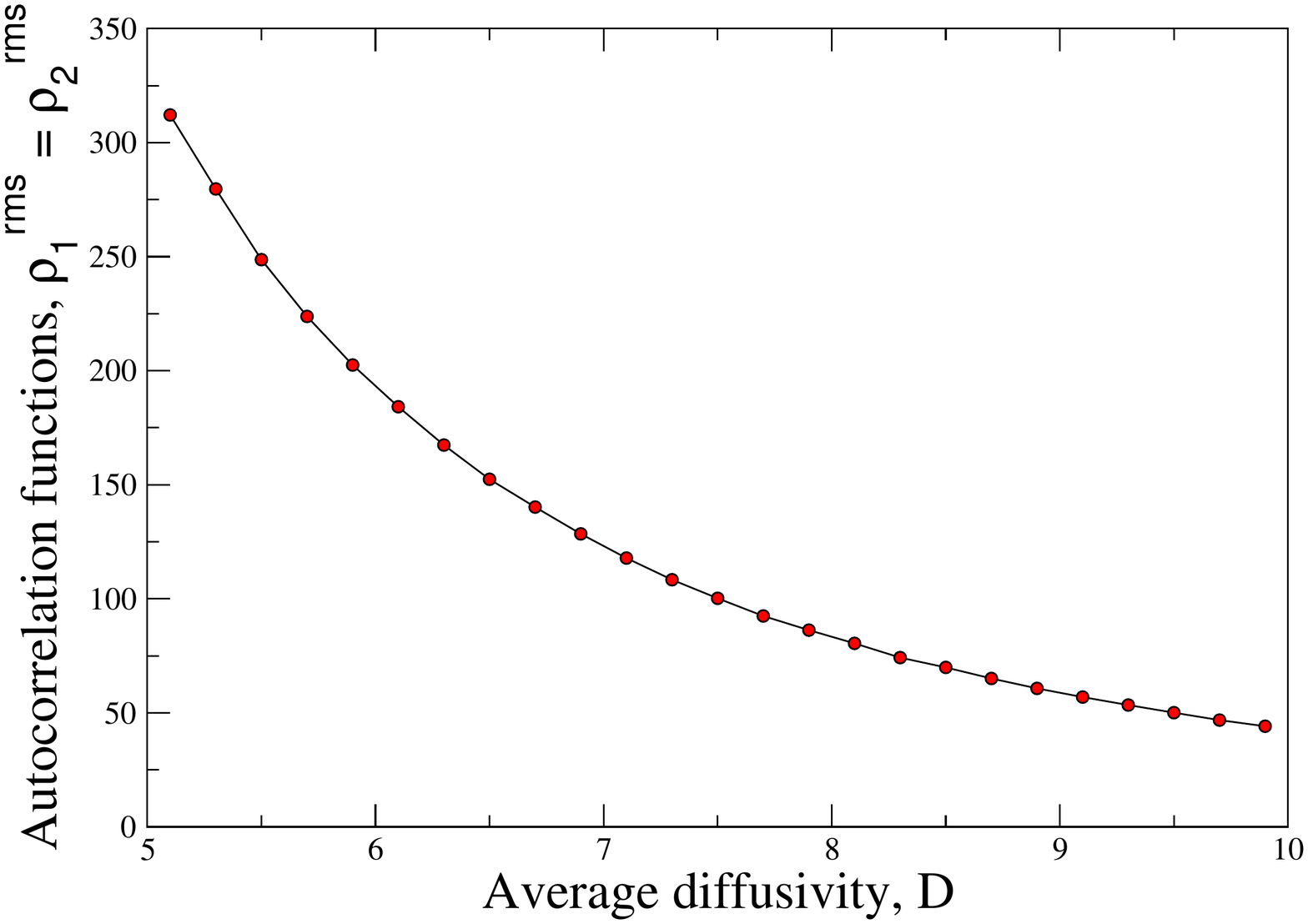}
\caption{Variation of the autocorrelation function $\rho_1^{\text{rms}}$ or $\rho_2^{\text{rms}}$ against the average scaled diffusivity $D$ (assumption $D_1=D_2=D$, scaling factor ${10}^{-10}$). The dots represent the actual data points obtained from a numerical solution of Eq. (\ref{auto}) for the parameter set $\kappa_1=1, \kappa_2=0.001$ for identical noise strengths $\gamma_1=\gamma_2=1$ (coefficient values are all scaled dimensionless numbers, based on \cite{Konstantinidis1998}). Results indicate a monotonic decay with $D$.\label{fig_autocorr}}
\end{figure}
\end{center}

For $\gamma_1=\gamma_2$, as shown in Figure \ref{fig_autocorr}, the $\rho_1$ and $\rho_2$ plots merge with each other.

The following section alludes to the derivation of the spectral dynamics in the Fourier transformed space that could be easily encapsulated with the descriptions of the autocorrelation and cross-correlation functions, the former representing the \enquote{self-energy} structure while the latter relates to the \enquote{spatiotemporal} interaction of sites. 

In order to solve for both auto and cross-correlation functions, we will need to evaluate Eqs. (\ref{autocorr}) and (\ref{crosscorr}) around the poles which are defined through the eighth degree polynomial equation ${{\zeta_n}}(\omega)=0$, where the roots of the equation will be given by $\omega=\Omega_i$ ($i$ = 1, 2,..., 8). These poles are given in details in Appendix I. The nature of the complexity of the analytical structure can be gauged even from the simplified special case of $D_1=D_2$, represented by the equation 
${{\zeta_n}} ={(\omega+i\Omega_0)}^4 {(\omega-i\Omega_0)}^4= 0$,
where 

{{
\begin{equation}
\Omega_0 \approx \frac{1}{2^{1/4}}{\big[2D^2 k^4 + 4Dk^2(\kappa_1+\kappa_2)+2{(\kappa_1+\kappa_2)}^2\big]}^{1/4}.
\label{omegaeqn}
\end{equation}
}}

\begin{center}
\begin{figure}[tbp]
\includegraphics[height=10.0cm,width=12.0cm]{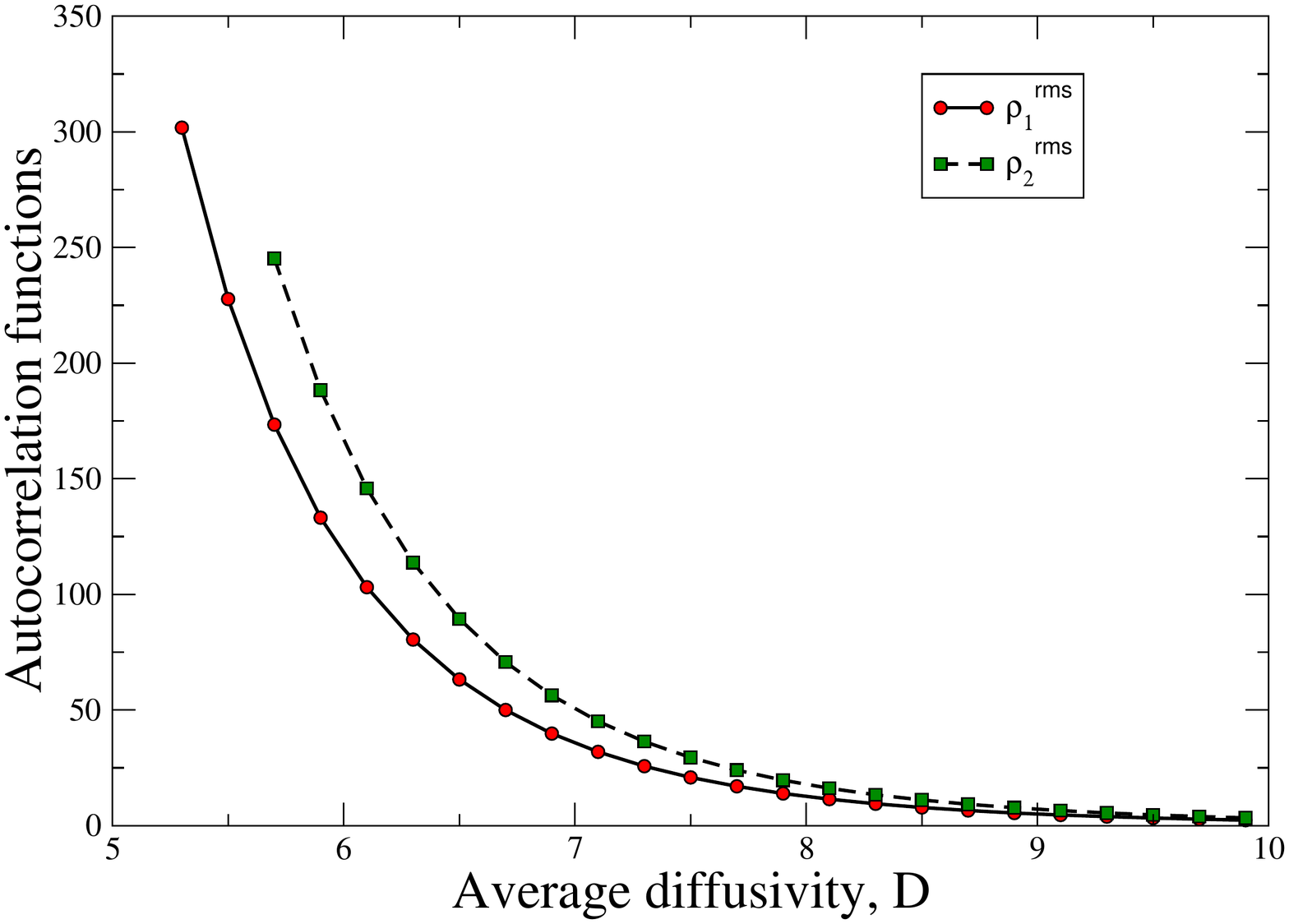}
\caption{Variation of the autocorrelation function $\rho_1^{\text{rms}}$ (represented by circles) or $\rho_2^{\text{rms}}$ (represented by squares) against the average scaled diffusivity $D$ (assumption $D_1=D_2=D$, scaling factor ${10}^{-10}$) for anisotropic noise case: $\gamma_1=1$ and $\gamma_2=2$. Results are obtained from a numerical solution of Eqs. (\ref{autorho}) and (\ref{auto}) respectively for $\rho_1^{\text{rms}}$ and $\rho_2^{\text{rms}}$ for $\kappa_1=1$ and $\kappa_2=0.001$. Coefficient values are all scaled dimensionless numbers, based on \cite{Konstantinidis1998}.
\label{fig_autocorrmixed}}
\end{figure}
\end{center}

\noindent
Within the ambits of this assumption ($D_1=D_2=D$), we arrive at the root-mean-squared autocorrelation functions for spatial dimension $d=3$ as follows

\begin{subequations}
\begin{eqnarray}
{\rho_1}^{\text{rms}} 
&=& \sqrt{2\pi^2 \int dk\:k^2\dfrac{\big[5\gamma_2 \kappa_2^2 + \gamma_1\big(5 D^2 k^4 + 
\Omega_0^4+5\kappa_2(2Dk^2+\kappa_2)\big)\big]}{8\Omega_0^7}}, 
\label{autorho}
\\
{\rho_2}^{\text{rms}} 
&=& \sqrt{2\pi^2 \int dk\:k^2\dfrac{\big[5\gamma_1 \kappa_1^2 + \gamma_2\big(5 D^2 k^4 + \Omega_0^4+5\kappa_1(2Dk^2+\kappa_1)\big)\big]}{8\Omega_0^7}}
\label{auto}
\end{eqnarray}
\label{autooo}
\end{subequations}

It is interesting to note that for $D_1=D_2$, the deterministic versions of Eqs. (\ref{modeleqna}) and (\ref{modeleqnb}) involving $\rho_1,\:\:\rho_2$ obey Eq. (\ref{scaledmodeleqn}) for a special form of the constants while $\rho_1+\rho_2$ still abides by Fick's law; however, if $D_1 \neq D_2$, then both $\rho_1,\:\rho_2$, and also $\rho_1+\rho_2$ obey Eq. (\ref{scaledmodeleqn}).  
The answer lies in the average energy dissipation rates of each of the two phases (IC and TJ) in the thermally driven system; these rates are equal to each other with exact values calibrated against the system parameters ($D,\kappa_1,\kappa_2$) involved. We should emphasize here that the simplifying assumption ($D_1=D_2=D$) used in arriving at the above result in no way sullies the implication of this analysis. For all realistic experimental observations \cite{Konstantinidis1998, Konstantinidis1999} concerning double diffusivity, including its application in explaining the oxygen diffusivity in barium superconductors \cite{Konstantinidis2001}, the two diffusive constants typically differ by about three orders of magnitude whose exact correlation forms can be analyzed using the representations in the Appendix. Figure \ref{fig_autocorr} uses identical noise strengths $\gamma_1=\gamma_2=1$. 

For anisotropic noise, the equivalent representation is provided in Fig. \ref{fig_autocorrmixed}. 

\subsection{Spatial Correlation of Phases}

\noindent
In this section, our attention will be focused on evaluating how the concentration of each phase changes with spatial distance in the dynamical equilibrium limit for spatial dimension $d=3$. Mathematically, this implies evaluation of the respective spatial correlation functions of each phase for all times and then taking ensemble averages over all noise realizations. 

By definition, we have

\begin{eqnarray}
{[{\rho_i^{\text{rms}}}({\bf r}) ]}^2 &=& \langle \rho_i({\bf x},t)*\rho_i({\bf x +r},t) \rangle \nonumber \\
&=& \int d^3{\bf k} \int d{\omega} e^{-i{\bf k}.{\bf r}}\:\langle \rho_i({\bf k},\omega)\:\rho_i^{*}(-{\bf k}-\omega) \rangle \nonumber \\
&=& 2\pi^2 \int dk \:k J_0(kr)\:\int d{\omega}\:\langle \rho_i({\bf k},\omega)\:\rho_i^{*}(-{\bf k}-\omega)\rangle \nonumber \\
&=& 8 \pi^3 D \int dk\: k J_0(kr)\dfrac{\big[5\gamma_2 \kappa_2^2 + \gamma_1\big(5 D^2 k^4 + \Omega_0^4+5\kappa_2(2Dk^2+\kappa_2)\big)\big]}{8\Omega_0^7},
\label{spatialphaseeqn1}
\end{eqnarray}

\noindent
where $J_0(x)$ represents the zeroth order Bessel function for the scalar variable $x$ and $i$=1, 2. The above Eq. (\ref{spatialphaseeqn1}) leads to the following density expression for both phases as given below

\begin{subequations}
\begin{eqnarray}
{\rho_1^{\text{rms}}}({\bf r}) &=& \sqrt{8 \pi^3 D \int_{k_0}^{k_m} dk\: k J_0(kr)\dfrac{\big[5\gamma_2 \kappa_2^2 + \gamma_1\big(5 D^2 k^4 + \Omega_0^4+5\kappa_2(2Dk^2+\kappa_2)\big)\big]}{8\Omega_0^7}},
\label{spatialphaseeqn2} \\
{\rho_2^{\text{rms}}}({\bf r}) &=& \sqrt{8 \pi^3 D \int_{k_0}^{k_m} dk\: k J_0(kr)\dfrac{\big[5\gamma_1 \kappa_1^2 + \gamma_2\big(5 D^2 k^4 + \Omega_0^4+5\kappa_1(2Dk^2+\kappa_1)\big)\big]}{8\Omega_0^7}},
\label{spatialphaseeqn22}
\end{eqnarray}
\end{subequations}

where $k_0$ and $k_m$ refer to the minimum and maximum of the wave vector $k$ (measured as the inverse of the characteristic system length). In our numerical solution, as shown in Fig. \ref{fig_spatcorr}, we have chosen $k_0=0$ and $k_m=10000$. {{The plots shown in Fig. 3 remain mostly unaffected by the specific choice of $k_0$ and $k_m$. This can be easily seen from a cross-check of the integration kernels given in Eqs. (\ref{spatialphaseeqn2}) and (\ref{spatialphaseeqn22}) in the limit $k\to$ large, as given below:
 \begin{center}
\begin{figure}[tbp]
\includegraphics[height=6.0cm,width=12cm]{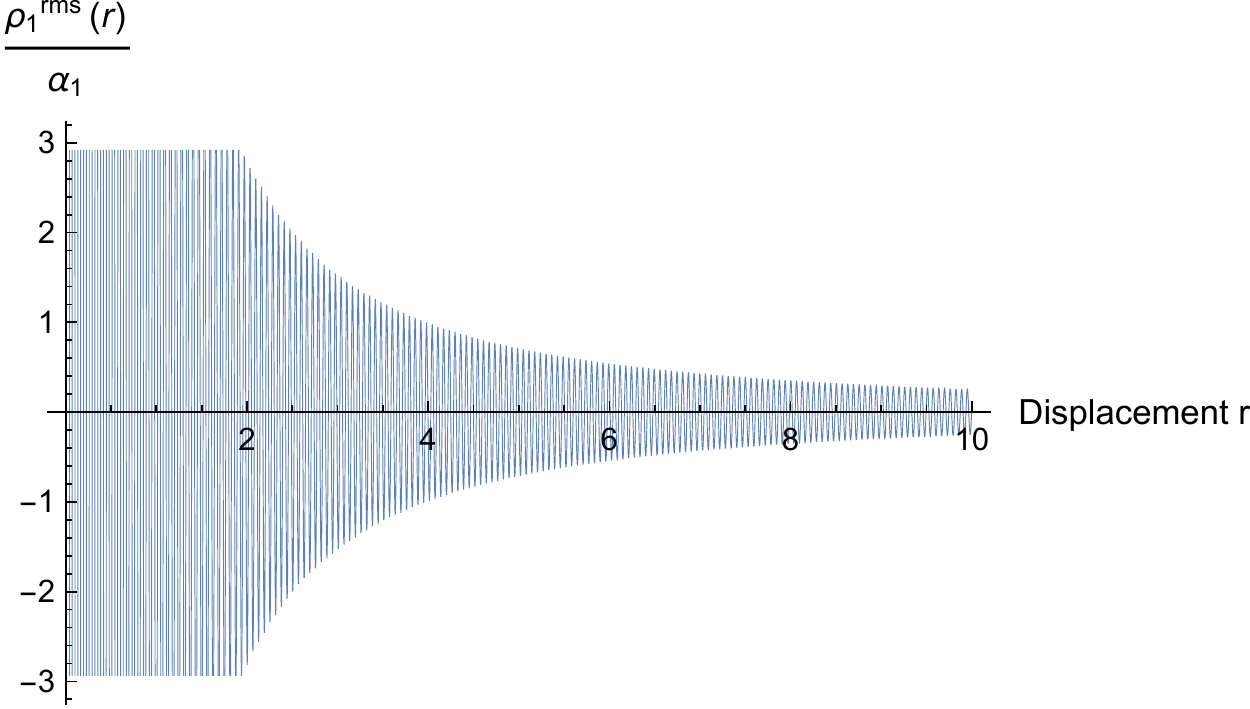}
\caption{Variation of the spatial correlation function $C_r=\rho_1^{\text{rms}}(r)=\rho_2^{\text{rms}}(r)$ against displacement $r$ for large $k$ (assumption $D_1=D_2=D={10}^{-9}$). The plot is obtained from a numerical solution of Eqs. (\ref{limitphaseeqn2}) and (\ref{limitphaseeqn22}) for the parameter rescaled version ({{where $\alpha_1$}}=$-\dfrac{\pi^2 (5 \gamma_2 \kappa_2^2+\gamma_1(5\kappa_2^2+2{(\kappa_1+\kappa_2)}^8))}{8\times 2^{3/4}{(\kappa_1+\kappa_2)}^{14}}$) for identical noise strengths $\gamma_1=\gamma_2=1$ (coefficient values are all scaled dimensionless numbers, based on \cite{Konstantinidis1998}). 
\label{figg4}}
\end{figure}
\end{center}

\begin{subequations}
\begin{eqnarray}
{{\rho_1^{\text{rms}}}({\bf r}) }_{k \to \text{large}} &=& \sqrt{8 \pi^3 D \int_{k_0}^{k_m} dk\: k J_0(kr) \bigg[|\dfrac{\pi^2 (5 \gamma_2 \kappa_2^2+\gamma_1(5\kappa_2^2+2{(\kappa_1+\kappa_2)}^8))}{8\times 2^{3/4}{(\kappa_1+\kappa_2)}^{14}}|\bigg]},
\label{limitphaseeqn2} \\
{{\rho_2^{\text{rms}}}({\bf r}) }_{k \to \text{large}} &=& \sqrt{8 \pi^3 D \int_{k_0}^{k_m} dk\: k J_0(kr) \bigg[|\dfrac{\pi^2 (5 \gamma_1 \kappa_1^2+\gamma_2(5\kappa_1^2+2{(\kappa_1+\kappa_2)}^8))}{8\times 2^{3/4}{(\kappa_1+\kappa_2)}^{14}}|\bigg]}.
\label{limitphaseeqn22}
\end{eqnarray}
\end{subequations}
Within a very short interval, contributions to the correlation functions from such large-spatial separations can be seen to decay to zero as shown in Fig. \ref{figg4}. This confirms the convergence of the integrals in Eqs. (\ref{spatialphaseeqn2}) and (\ref{spatialphaseeqn22}). Figure \ref{figg4} validates the convergence of Eq. (\ref{spatialphaseeqn2}); a similar analysis could be repeated for Eq. (\ref{spatialphaseeqn22}) in the limit of $k\rightarrow$ large to establish a similar convergence.}}
It must be remembered, though, that $\rho_1^{\text{rms}}=\rho_2^{\text{rms}}$ is a result of our assumption $D_1=D_2$; the correlations will have different values for the different phases for $D_1 \neq D_2$, the precise nature of which can be estimated from the expressions presented in the Appendix. Once again, for the isotropic case ($\gamma_1=\gamma_2$), as shown in Fig. \ref{fig_spatcorr}, the $\rho_1$ and $\rho_2$ plots merge with each other.

In order to get a feel for the functional dependence of the spatial correlation function for the \enquote{nondegenerate} case $D_1 \neq D_2$, the relevant correlation functions can be approximately shown to be as follows (details in the Appendix)
\begin{center}
\begin{figure}[tbp]
\includegraphics[height=10.0cm,width=12cm]{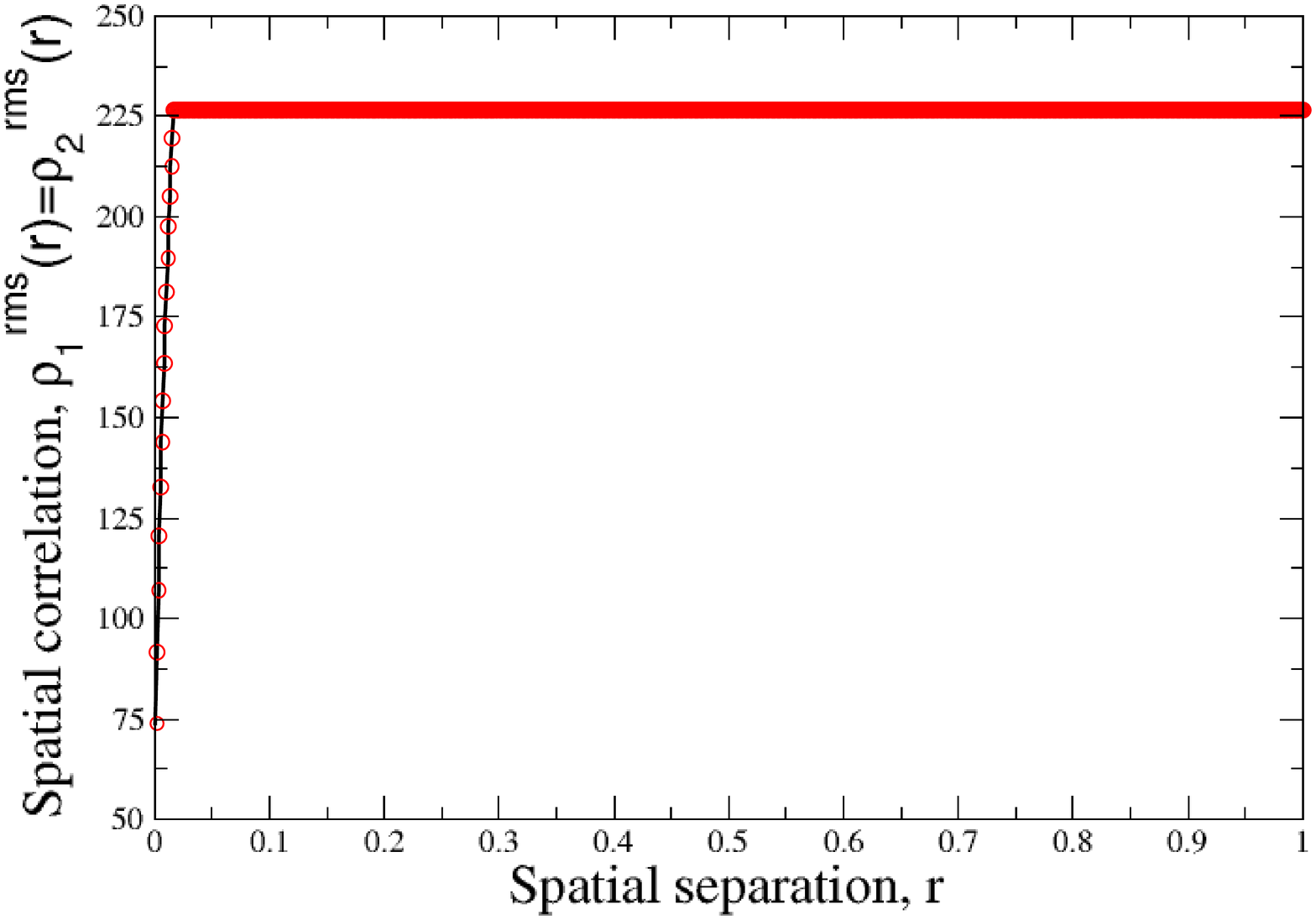}
\caption{Variation of the spatial correlation function $C_r=\rho_1^{\text{rms}}(r)=\rho_2^{\text{rms}}(r)$ against displacement $r$ (assumption $D_1=D_2=D={10}^{-9}$). The plot is obtained from a numerical solution of Eqs. (\ref{spatialphaseeqn2}) and (\ref{spatialphaseeqn22}) for the parameter set $\kappa_1=1, \kappa_2=0.001$ for identical noise strengths $\gamma_1=\gamma_2=1$ (coefficient values are all scaled dimensionless numbers, based on \cite{Konstantinidis1998}). {{The circles represent the real data points (the saturation region is represented by multiple close-lying circles doubling up as a thick solid straight line) while the solid straight line is the extrapolated fit.}}
\label{fig_spatcorr}}
\end{figure}
\end{center}

\begin{center}
\begin{figure}[tbp]
\includegraphics[height=10.0cm,width=12.0cm]{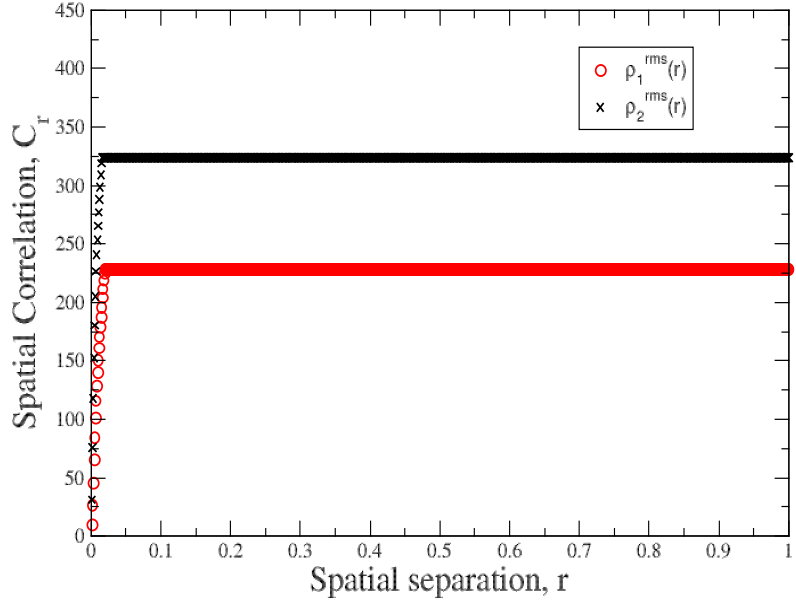}
\caption{Variation of the spatial correlation function $C_r$ for the variables {{$\rho_1^{\text{rms}}(r)$ (represented by circles) or $\rho_2^{\text{rms}}(r)$ (represented by crosses)}} against the separation distance $r$ (assumption $D_1=D_2=D={10}^{-9}$) for the anisiotropic noise case: $\gamma_1=1$ and $\gamma_2=2$. Results are obtained from a numerical solution of Eqs. (\ref{spatialphaseeqn2}) and (\ref{spatialphaseeqn22}) for $\kappa_1=1$ and $\kappa_2=0.001$. Coefficient values are all scaled dimensionless numbers, based on \cite{Konstantinidis1998}. {{The saturation regime for both plots are represented by multiple close lying symbols, circles or crosses, as the case may be, giving them the appearance of a thick line.}}
 \label{fig_spatcorrmixed}}
\end{figure}
\end{center}

{{
\begin{center}
\begin{figure}[tbp]
\includegraphics[height=10.0cm,width=12cm]{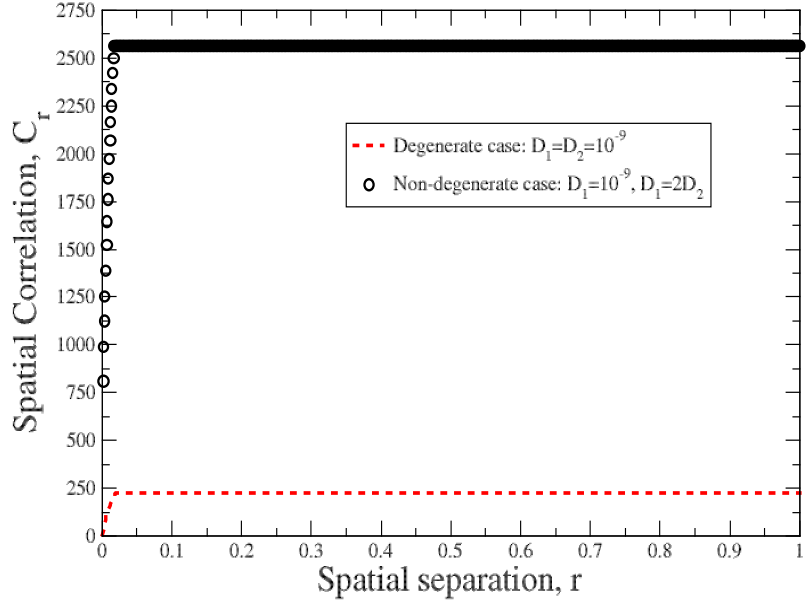}
\caption{Variation of the spatial correlation functions $C_r=\rho_1^{\text{rms}}(r)$ against displacement $r$ for the {{degenerate case ($D_1=D_2={10}^{-9}$; represented by dots) versus the non-degenerate case ($D_1=2D_2$; $D_2={10}^{-9}$; represented by circles)}}. The plot is obtained from a numerical solution of Eqs. (\ref{nondegphaseeqn2}) and (\ref{nondegphaseeqn22}) for the parameter set $\kappa_1=1, \kappa_2=0.001$ for identical noise strengths $\gamma_1=\gamma_2=1$ (coefficient values are all scaled dimensionless numbers, based on \cite{Konstantinidis1998}). 
\label{fig_spatcorrcomp}}
\end{figure}
\end{center}

\begin{subequations}
\begin{eqnarray}
{\rho_1^{\text{rms}}}({\bf r}) &\approx& \sqrt{8 \pi^3 D_{\text{av}} \int_{k_0}^{k_m} dk\: k J_0(kr)\dfrac{\big[5\gamma_2 \kappa_2^2 + \gamma_1\big(5 D_{\text{av}}^2 k^4 + \Omega_{\text{av}}^4+5\kappa_2(2D_{\text{av}}k^2+\kappa_2)\big)\big]}{8\Omega_{\text{av}}^7}},
\label{nondegphaseeqn2} \\
{\rho_2^{\text{rms}}}({\bf r}) &\approx& \sqrt{8 \pi^3 D_{\text{av}} \int_{k_0}^{k_m} dk\: k J_0(kr)\dfrac{\big[5\gamma_1 \kappa_1^2 + \gamma_2\big(5 D_{\text{av}}^2 k^4 + \Omega_{\text{av}}^4+5\kappa_1(2D_{\text{av}}k^2+\kappa_1)\big)\big]}{8\Omega_{\text{av}}^7}},
\label{nondegphaseeqn22}
\end{eqnarray}
\end{subequations}
where $D_{\text{av}}=\frac{D_1+D_2}{2}$ and $\Omega_{\text{av}}= -\dfrac{\bigg{(-T_1-k^2 D_1(2\kappa_1+T_2) -(\kappa_1+\kappa_2)(\kappa_1+\kappa_2+T_2)-k^2 D_2(2\kappa_2+T_2) \bigg)}^{1/4}}{2^{1/4}}$, with $T_1=k^4 (D_1^2+D_2^2)$ and $T_2=\sqrt{{\bigg(k^2(D_1-D_2)+\kappa_1\bigg)}^2+2\bigg(k^2(D_2-D_1)+\kappa_1\bigg)\kappa_2+\kappa_2^2}$.

For the special case of $\gamma_1=\gamma_2=1$, comparing with the \enquote{degenerate} case $D_1=D_2={10}^{-9}$, we arrive at a very similar functional behavior for the non-degenerate case as well.
}}

In Figure \ref{fig_spatcorr}, we show how the spatial correlation saturates with increasing separation distance, for the case $\gamma_1=\gamma_2$. In order to portray the situation for the case of anisotropic noise ($\gamma_1 \neq \gamma_2$), below we plot Fig. \ref{fig_spatcorrmixed}. While the qualitative features remain unchanged, due to a large $\gamma_2=2\gamma_1$, the saturation level of $\rho_2^{\text{rms}}$ can be seen to be way above $\rho_1^{\text{rms}}$, although the crossover point remains roughly unchanged. This feature clearly suggests that noise anisotropy is not a qualitatively devolving feature of this dynamics.

Unlike the deterministic case (as in \cite{Aifantis2011}), the dynamical equilibrium of the thermal noise driven two-phased system stabilizes to the same spatial concentration spread for both phases. This clearly suggests a difference at the qualitative level, as well as obvious quantitative differences (compared to \cite{Aifantis2011}).

\subsection{Temporal correlation of phases}

It is well known that a fundamental consideration in multi-phase systems is the time evolution of the interface separating two different phases. Often such systems are known to be stochastically perturbed and hence nonequilibrium in nature, potentially rendering the relevant dynamics as oscillatory, or with oscillatory-rotatory instability leading to chaos \cite{Hu2003}. Phase control through synchronization driving such systems away from the chaotic bifurcation point has in fact benchmarked the {\it hare-lynx} model in ecology \cite{Blasius2000}.  

By analogy with two-phase systems, we calculate below the
theoretical quantities which, in principle, can be compared with the
experimental set ups describing the temporal correlation dynamics. In the present case, the relevant variables necessary to model such a stochastically driven two-phase system are ${{\rho_i}^{\text{rms}}(T)}$, which are encapsulated in the following equations

\begin{eqnarray}
{[{\rho_i}^{\text{rms}}(T)]}^2 &=& \langle \rho_i({\bf x},t)*\rho_i({\bf x},t+T)\rangle \nonumber \\
&=& \int d^3{\bf k} \int d{\omega} e^{-i\omega T}\:\langle \rho_i({\bf k},\omega)\:\rho_i^{*}(-{\bf k}-\omega)\rangle,
\end{eqnarray}
for $i=1,2$. Our next target is to estimate the two-point temporal correlation functions, as the first-order euphemism of the probability density function (also connected to the fluctuation-dissipation theorem \cite{Barabasi}). 
\begin{center}
\begin{figure}[tbp]
\includegraphics[height=10.0cm,width=12.0cm]{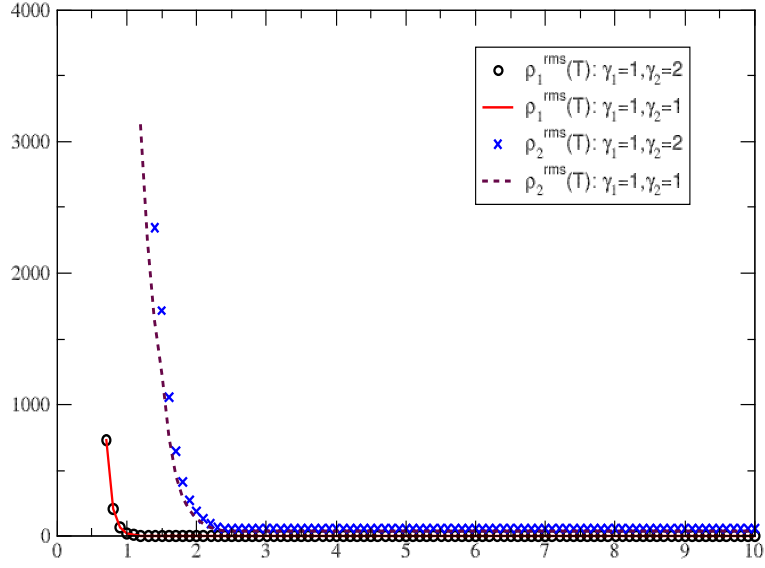}
\caption{Variation of the temporal correlation function $C_T$ for the variables $\rho_1(T)$ and $\rho_2(T)$ against the time difference $T$ (assumption $D_1=D_2=D={10}^{-9}$). {{The circles and crosses respectively represent $\rho_1^{\text{rms}}(T)$ and $\rho_2^{\text{rms}}(T)$ for the special case $\gamma_1=\gamma_2=1$ while the corresponding anisotropic noise cases ($\gamma_1=1,\:\gamma_2=2$) are represented by the solid line and dots respectively}}. Results are obtained from a numerical solution of Eqs. (\ref{temporalphaseeqn2}) and (\ref{temporalphaseeqn22}) for $\kappa_1=1$ and $\kappa_2=0.001$. Coefficient values are all scaled dimensionless numbers, based on \cite{Konstantinidis1998, Konstantinidis1999}.
\label{fig_tempcorrmixed}}
\end{figure}
\end{center}
Calculating as before, we obtain these two-point correlation functions as follows
\begin{subequations}
\begin{eqnarray}
{\big[\rho_1^{\text{rms}}(T)\big]}^2 &=& \dfrac{\gamma_1 \pi^3}{3}\int\:dk \:\bigg(\dfrac{k^2}{\Omega_0^7}\bigg)\:e^{-T\Omega_0}\bigg[ \big(15+T\Omega_0(15+T\Omega_0(6+T\Omega_0))\big)\gamma_2 \kappa_2^2 \nonumber \\
&+& \gamma_1\big(\Omega_0^4(3+T\Omega_0(3+T\Omega_0(-6+T\Omega_0))\big) \nonumber \\
&+& D^2 k^4\big(15+T\Omega_0(15+T\Omega_0(6+T\Omega_0))\big) \nonumber \\
&+&  \kappa_2\big(15+T\Omega_0(15+T\Omega_0(6+T\Omega_0))\big) (2Dk^2 + \kappa_2)) \bigg]
\label{temporalphaseeqn2}
\end{eqnarray}
\begin{eqnarray}
{\big[\rho_2^{\text{rms}}(T)\big]}^2 &=& \dfrac{\gamma_2 \pi^3}{3}\int\:dk \:\bigg(\dfrac{k^2}{\Omega_0^7}\bigg)\:e^{-T\Omega_0}\bigg[ \big(15+T\Omega_0(15+T\Omega_0(6+T\Omega_0))\big)\gamma_1 \kappa_1^2 \nonumber \\
&+& \gamma_2\big(\Omega_0^4(3+T\Omega_0(3+T\Omega_0(-6+T\Omega_0))\big) \nonumber \\
&+& D^2 k^4\big(15+T\Omega_0(15+T\Omega_0(6+T\Omega_0))\big) \nonumber \\ &+& \kappa_1\big(15+T\Omega_0(15+T\Omega_0(6+T\Omega_0))\big) (2Dk^2 + \kappa_1)) \bigg]
\label{temporalphaseeqn22}
\end{eqnarray}
\end{subequations}

A remarkable feature of the temporal correlation function plot as shown in Fig. \ref{fig_tempcorrmixed} is the relatively low effect of the noise (an)isotropy compared to the spatial cases. The other aspect of these temporal correlation functions shown in Eqs. (\ref{temporalphaseeqn2}) and (\ref{temporalphaseeqn22}) is the tremendous stability with regard to noise fluctuations. In our simulations, we sampled across a wide range of noise strengths ${10}^{-2}<D<{10}^{-9}$ to find that the results shown in Fig. \ref{fig_tempcorrmixed} remain unaffected by the value of $D$, as long as the stability condition $k_{\text{min}}>\sqrt{\dfrac{D}{\lambda_2}}$ is obeyed, where $k_{\text{min}}$ is the minimum allowed value of the wave vector $k$.

In the analysis above, we have deliberately refrained from going into the details of the cross-correlation description, both for the spatial as well as the temporal cases. While qualitatively the presence of an imaginary part in the correlation function indicates attenuation, the subject will be dealt with in more detail separately.  

\section{Nanodiffusion Spatiotemporal Dynamics}

\noindent
In order to compare nanodiffusion transport between the present thermally driven model and the paradigmatic Aifantis model \cite{Aifantis2011}, we will now calculate the spatiotemporal dynamics of the concentration fields $\rho_1,\:\rho_2$ and $\rho=\rho_1+\rho_2$, based on Model 2, starting from the conserved SGNM model previously defined in Eq. (\ref{scaledmodeleqn}). The corresponding solution of the phase concentrations as shown in \cite{Aifantis2011} will be compared with its thermally driven counterpart for periodic boundary conditions. As always, in the context of stochastic Brownian-type statistics, individual dynamical variables give way to their corresponding r.m.s. counterparts. Starting from Eq. (\ref{scaledmodeleqn}) and using $\lambda_1=c^*D$ and $\lambda_2=cD$, we can rewrite the model as

\begin{equation}
\partial_t \rho + \tau {\partial_t}^2 \rho = D \nabla^2 \rho + \lambda_1 \partial_t\:\nabla^2 \rho + \lambda_2\:\nabla^4 \rho + \eta({\bf x},t).
\label{interfaceeqn}
\end{equation}

\noindent
Fourier transformation of the above Eq. (\ref{interfaceeqn}) in the ${\bf k}-\omega$ space gives
\begin{equation}
\hat \rho({\bf k},\omega) = \frac{\hat \eta({\bf k},\omega)}{-i\omega (1+\gamma_1 k^2) + (Dk^2 -\tau \omega^2 -\gamma_2 k^4)}.
\label{rhoeqn}
\end{equation}

The above Eq. (\ref{rhoeqn}) has 4 poles at $\omega=\omega_i$ ($i$ = 1, 2, 3, 4), such that $\omega_i=\pm \sqrt{T_1 \pm \dfrac{\sqrt{T_2}}{2\tau^2}}$, in which $T_1 = -\frac{1}{2\tau^2}+\frac{Dk^2}{\tau}-\frac{\lambda_1 k^2}{\tau^2}-\frac{\lambda_1^2 k^4}{2\tau^2}-\frac{\lambda_2 k^4}{\tau}$, $T_2 = {(1-2Dk^2 \tau+2\lambda_1 k^2 + \lambda_1^2 k^4+2k^4 \tau \lambda_2)}^2-4\tau^2(D^2 k^4 -2D\lambda_2 k^6+\lambda_2^2 k^8)$. To simplify calculations through a reduced model, we study the case for an {\it overdamped} system where $\tau \to 0$. 
\begin{center}
\begin{figure}[tbp]
\includegraphics[height=10.0cm,width=12.0cm]{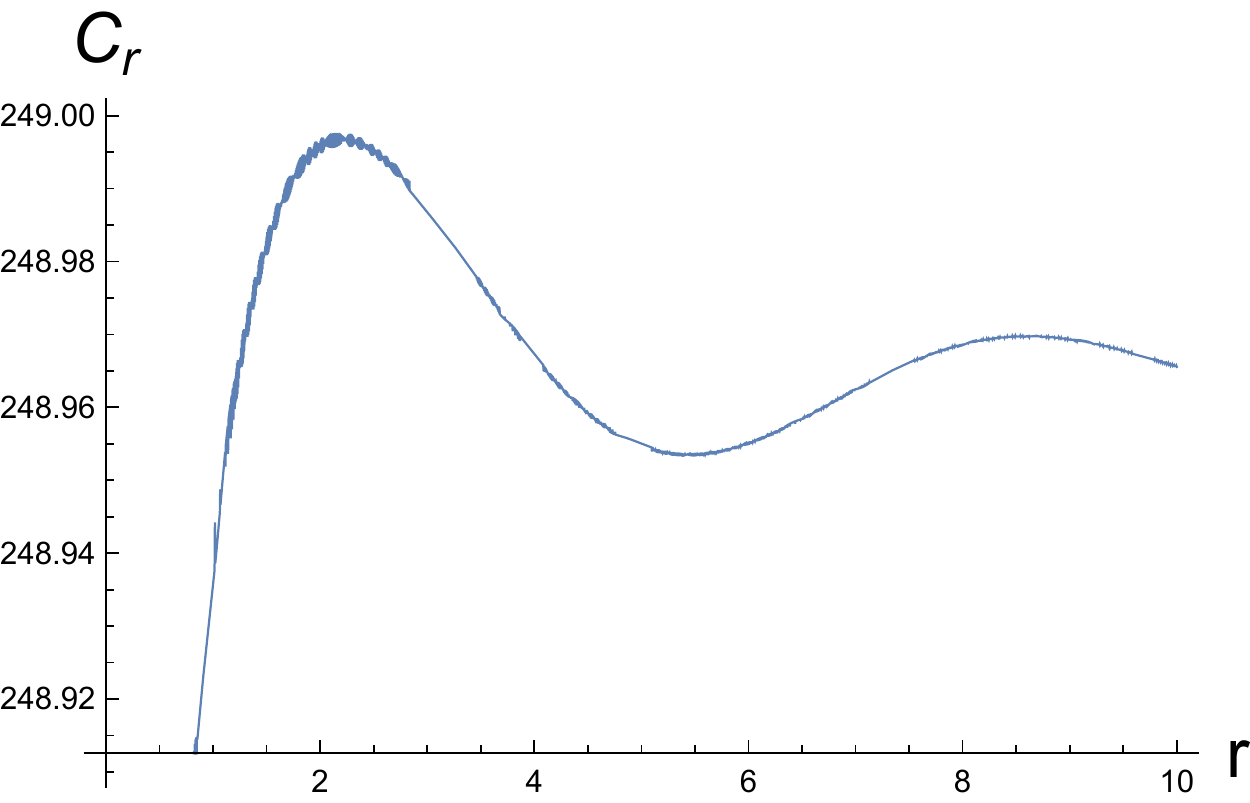}
\caption{Variation of the spectral correlation function $C_r$ against the separation distance $r$ (assumption $D_1=D_2=D={10}^{-1},\gamma_0=1$). Results are obtained from a numerical solution of Eq. (\ref{spatialmodel2}) for $\lambda_1=\lambda_2=1$.
\label{fig_nanospat}}
\end{figure}
\end{center}
\begin{center}
\begin{figure}[tbp]
\includegraphics[height=10.0cm,width=12.0cm]{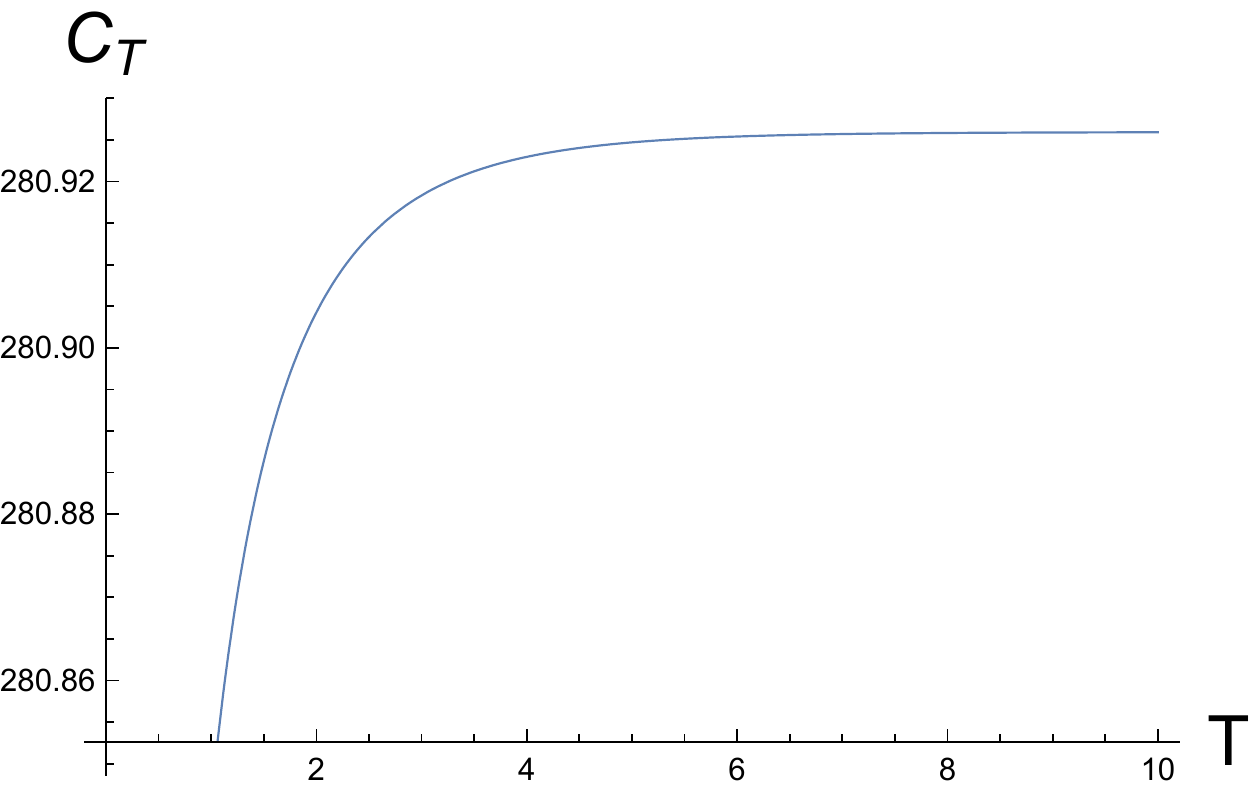}
\caption{Variation of the spectral correlation function $C_T$ against the time difference $T$ (assumption $D_1=D_2=D={10}^{-1}, \gamma_0=1$). Results are obtained from a numerical solution of Eq. (\ref{temporalmodel2}) for $\lambda_1=\lambda_2=1$.
\label{fig_nanotemp}}
\end{figure}
\end{center}
In this limit, the two poles turn out as $\pm i\Gamma$, where $\Gamma= \dfrac{k^2 (D-\lambda_2 k^2)}{1+\lambda_1 k^2}$. This gives autocorrelation function as
\begin{eqnarray}
\rho_{\text{auto}}^{\text{rms}} 
&=& \sqrt{2\pi^2 \gamma_0 \displaystyle \int_{\sqrt{\frac{D}{\lambda_2}}}^{\infty} \:\frac{k\sqrt{1+\lambda_1 k^2}}{\sqrt{Dk^2 -\lambda_2 k^4}}\:dk},
\label{rhormseqn}
\end{eqnarray}
where $\gamma_0$ is the strength of the Gaussian white noise.

A comparison between Eqs. (\ref{autorho}) and (\ref{auto}), and Eq. (\ref{rhormseqn}) allows us to compare the quantitative difference between individual concentrations with individually added noise as against their total concentration with an overall added noise from the perspective of energy dissipation rate. This will be separately evaluated through the {\it spatial autocorrelation function} and the {\it temporal autocorrelation function} [definitions as in equations (\ref{spatialphaseeqn1}), and (\ref{temporalphaseeqn2}), and (\ref{temporalphaseeqn22})] that are defined as per \cite{Amit1,Amit2,Amit3} as follows

\begin{subequations}
\begin{eqnarray}
C_r={\rho^{\text{rms}}}({\bf r}) &=&\displaystyle \sqrt{2\pi^3 \gamma_0\:\displaystyle \int_{\sqrt{\frac{D}{\lambda_2}}}^{\infty}\: dk\: \frac{k J_0(kr)\sqrt{1+\lambda_1 k^2}}{\sqrt{Dk^2-\lambda_2 k^4}}-\rho_{\text{auto}}^{\text{rms}}} 
\label{spatialmodel2} \\
C_T={\rho^{\text{rms}}}(T) &=& \displaystyle \sqrt{8\pi^2 \gamma_0\:\displaystyle \int_{\sqrt{\frac{D}{\lambda_2}}}^{\infty}\:dk \:e^{-T \Gamma({\bf k},\omega)} \frac{k \sqrt{1+\lambda_1 k^2}}{\sqrt{Dk^2-\lambda_2 k^4}}-\rho_{\text{auto}}^{\text{rms}}}
\label{temporalmodel2}
\end{eqnarray}
\end{subequations}

Figure \ref{fig_nanospat} shows a periodic stabilizing pattern which is distinctly different from either of Fig. \ref{fig_spatcorr} or Fig. \ref{fig_spatcorrmixed}. This is most remarkable since the deterministic description does not show any qualitative difference between the two phase model and its equivalent single phase description \cite{Konstantinidis1998}. The temporal description (Fig. \ref{fig_nanotemp}) too shows qualitatively distinctive features compared to Fig. \ref{fig_tempcorrmixed}. Fig. \ref{fig_nanotemp} indicates a sharp rise towards a saturation concentration density, a case of finite-sized saturation effect, as opposed to the decaying pattern represented in the anisotropic case in Fig. \ref{fig_tempcorrmixed}. {{In order to magnify the quantitative impact, we used a larger noise value $D=10^{-1}$ in arriving at these plots; however, the choice of the actual value does not impact the qualitative feature in any way, as the model remains remarkably stable to noise perturbations [detailed earlier after Eqs. (19a) and (19b)]. In fact, the quantitative corrections are less than 1\% with every order change in noise [using formulas in Eqs. (\ref{spatialmodel2}) and (\ref{temporalmodel2})], reconfirming the noise amplitude independence of the dynamics. Here we must indicate, though, that while the dynamics is largely unaffected by changes in the noise amplitude, the noise distribution function is expected to be vitally important to the dynamics, a feature that is presently being studied for future publications.}}

This {\it reduced} model has a cut-off at $k=\sqrt{\frac{D}{\lambda_2}}$ which defines its validity regime. We postpone analysis of the full model, includinga  non-zero $\tau$, as defined in Eq. (\ref{interfaceeqn}) for a later work.

\section{Conclusion}
\label{section_Conclusion}
In this article, we have provided an initial analysis of the continuum double diffusivity model under stochastic forcing. Stochasticity is introduced separately as a white thermal noise, either in a Fick class of equations  describing the concentrations in each type of diffusion path, or in a higher-order equation for the total concentration. Comparisons between the two cases have been made by comparing the respective classes of spatiotemporal cross-correlations and autocorrelations. 

The autocorrelation plots in Figs. 1 and 2 conform qualitatively to predictions based on the deterministic double diffusivity models \cite{Konstantinidis1998, Konstantinidis1999, Konstantinidis2001}; although the amplitudes are higher due to additional energy inputs through stochastic forcing. It is pertinent to remember that the root-mean-squared forms of the respective autocorrelation forms are the dimensional equivalents of the corresponding quantities in the deterministic models in \cite{Aifantis2016,Konstantinidis1998, Konstantinidis1999, Konstantinidis2001} and hence could be compared on a term-by-term basis. As to the crosscorrelation terms in the spatiotemporal dynamics (Figs. 3 to 7), the stochastic model ushers in a new regime of description where stochasticity mediates off-diagonal, often asymmetric forcing across multiple variables even at the first Gaussian approximation order. While this is very much an expected part of real life nanodynamic processes, the deterministic double-diffusivity model failed to capture this aspect that we have successfully made now. The results are verifiable using experimental data.

An important aspect of this analysis is the relative independence of both spatial and temporal correlation functions to the stochastic fluctuations. Over a wide range of noise strengths (${10}^{-9} < D < {10}^{-2}$), the correlation functions showed no qualitative change and very little quantitative change, thereby confirming the stability of this model to noise perturbations. This indirectly explains why some past theories \cite{Konstantinidis1998, Konstantinidis1999, Konstantinidis2001} have managed to arrive at experimental results reasonably accurately for some cases while faltering in others.

In summary, we point out some interesting features that stochasticity brings into the double diffusivity model, comparing existing deterministic terms against the relative stochastic forcing. This is only a first step toward integrating the double diffusivity properties with real life fluctuations that could modulate the process. As this model and the corresponding higher-order diffusion equation have been shown to effectively interpret transport in heterogeneous media possessing more than one family of conduction paths, as well as transport phenomena at the nanoscale, more detailed analysis focusing on the precise nature of randomness, e.g., forcing through non-Gaussian noise, as well as also delving deeper into the microscopic dynamics of the process, will be pursued in the next set of publications.   

\section{Acknowledgments}
The combined support of ERC (Belgium)-13 and ARISTEIA II projects funded by GSRT (Greece) of the Green Ministry of Education, sponsoring A.K.C's visits to Thessaloniki, is gratefully acknowledged. 

\section{Appendix I: \\ Poles of the autocorrelation function}

The poles of the autocorrelation function as defined in Eq. (\ref{autocorr}) can be obtained as follows:

\begin{subequations}
\begin{eqnarray}
\Omega_1 &=& -\dfrac{\bigg{(-T_1-k^2 D_1(2\kappa_1+T_2) -(\kappa_1+\kappa_2)(\kappa_1+\kappa_2+T_2)-k^2 D_2(2\kappa_2+T_2) \bigg)}^{1/4}}{2^{1/4}}, \\
\Omega_2 &=& -i\dfrac{\bigg{(-T_1-k^2 D_1(2\kappa_1+T_2) -(\kappa_1+\kappa_2)(\kappa_1+\kappa_2+T_2)-k^2 D_2(2\kappa_2+T_2) \bigg)}^{1/4}}{2^{1/4}}, \\
\Omega_3 &=& i\dfrac{\bigg{(-T_1-k^2 D_1(2\kappa_1+T_2) -(\kappa_1+\kappa_2)(\kappa_1+\kappa_2+T_2)-k^2 D_2(2\kappa_2+T_2) \bigg)}^{1/4}}{2^{1/4}}, \\
\Omega_4 &=& \dfrac{\bigg{(-T_1-k^2 D_1(2\kappa_1+T_2) -(\kappa_1+\kappa_2)(\kappa_1+\kappa_2+T_2)-k^2 D_2(2\kappa_2+T_2) \bigg)}^{1/4}}{2^{1/4}}, \\
\Omega_5 &=& -\dfrac{\bigg{(-T_1+k^2 D_1(-2\kappa_1+T_2) +(\kappa_1+\kappa_2)(-\kappa_1-\kappa_2+T_2)+k^2 D_2(-2\kappa_2+T_2) \bigg)}^{1/4}}{2^{1/4}}, \\
\Omega_6 &=& -i\dfrac{\bigg{(-T_1+k^2 D_1(-2\kappa_1+T_2) +(\kappa_1+\kappa_2)(-\kappa_1-\kappa_2+T_2)+k^2 D_2(-2\kappa_2+T_2) \bigg)}^{1/4}}{2^{1/4}}, \\
\Omega_7 &=& i\dfrac{\bigg{(-T_1+k^2 D_1(-2\kappa_1+T_2) +(\kappa_1+\kappa_2)(-\kappa_1-\kappa_2+T_2)+k^2 D_2(-2\kappa_2+T_2) \bigg)}^{1/4}}{2^{1/4}}, \\
\Omega_8 &=& \dfrac{\bigg{(-T_1+k^2 D_1(-2\kappa_1+T_2) +(\kappa_1+\kappa_2)(-\kappa_1-\kappa_2+T_2)+k^2 D_2(-2\kappa_2+T_2) \bigg)}^{1/4}}{2^{1/4}},
\end{eqnarray}
\end{subequations}

where $T_1=k^4 (D_1^2+D_2^2)$ and $T_2=\sqrt{{\bigg(k^2(D_1-D_2)+\kappa_1\bigg)}^2+2\bigg(k^2(D_2-D_1)+\kappa_1\bigg)\kappa_2+\kappa_2^2}$.

\end{document}